\documentclass[12pt]{iopart}
\usepackage{iopams,epsf,graphicx}
\begin{document}

\title{Nernst quantum oscillations in bulk semi-metals}

\author{Zengwei Zhu$^{1}$, Huan Yang\footnote{Current address: Institute of Physics and National Laboratory for Condensed Matter Physics, Chinese Academy of Sciences, China}$^{1}$, Aritra Banerjee\footnote{Current address: Department of Physics, University of Calcutta, Kolkata, India}$^{1}$, Liam Malone$^{2}$, Beno\^it Fauqu\'e$^{1}$ and Kamran Behnia$^{1}$}

\address{(1)LPEM (UPMC-CNRS), Ecole Sup\'erieure de Physique et de Chimie Industrielles, Paris, France\\
(2)Laboratoire National des Champs Magn\'{e}tiques Intenses (CNRS), Grenoble, France}
\ead{kamran.behnia@espci.fr}
\begin{abstract}
 With a widely available magnetic field of 10 T, one can attain the quantum limit in bismuth and graphite. At zero magnetic field, these two elemental semi-metals host a dilute liquid of carriers of both signs.  All quasi-particles are confined to a few Landau tubes, when the quantum limit is attained. Each time a Landau tube is squeezed before definitely leaving the Fermi surface,  the Nernst response sharply peaks. In bismuth, additional Nernst peaks, unexpected in the non-interacting picture, are resolved beyond the quantum limit. The amplitude of these unexpected Nernst peaks become more pronounced in the samples with the longest electron mean-free-path.
\end{abstract}

\maketitle

\section{Introduction}

During the last decade, following the observation of an anomalous Nernst signal in underdoped cuprates\cite{xu}, the Nernst effect has been used as a very sensitive probe of electron organization in solids\cite{behnia1}. Besides being a probe of superconducting fluctuations in the normal state of  cuprate\cite{wang}, organic\cite{nam} or conventional\cite{pourret} superconductors, the Nernst effect has been found to be significantly more sensitive than other transport coefficients to Fermi surface reconstruction in YBa$_{2}$Cu$_{3}$O$_{7-\delta}$\cite{cyr}, to quantum criticality in CeCoIn$_5$\cite{izawa} or to the onset of the hidden-order phase transition in URu$_2$Si$_2$\cite{bel}.

Giant quantum oscillations of the Nernst effect in two bulk elemental semi-metals, bismuth\cite{behnia2} and graphite\cite{zhu} were discovered in such a context. Both these systems were subject to intense experimental and theoretical exploration for several decades. Quantum oscillations of their thermoelectric response were observed and reported a long time ago\cite{steele,mangez,woollam}. However, only very recent studies were extended to higher fields and lower temperatures and resolved the particular profile of these oscillations in the Nernst channel as opposed to the Seebeck channel. The quantum oscillations of the Nernst response easily dwarf those seen in other transport coefficients. Moreover, the Nernst [and not the Seebeck] signal sharply peaks whenever a Landau level intersects the chemical potential. These features are absent in two-dimensional systems pointing to a link between dimensionality and the Nernst response\cite{zhu}.

\section{Quantum limit and its accessibility in bulk systems}

The quantum limit is attained when the magnetic field is strong enough to confine electrons to their lowest Landau level(s). In this limit, the cyclotron energy, $\hbar\omega_{c}$ where $\omega_{c}=\frac{eB}{m^{*}}$ is the cyclotron frequency, becomes comparable to the Fermi Energy, $\epsilon_F$. For ordinary bulk metals, the magnetic field necessary to attain this limit is in the several hundred Tesla range, well beyond the limits of current technology. In metals host to correlated electrons, the magnitude of the Fermi energy is significantly reduced by mass renormalization, but this does not suffice to make the the quantum limit accessible, since the same is true for the cyclotron energy.

In an equivalent  statement, the quantum limit is attained when the magnetic length, $\ell_{B}= \sqrt{\hbar/e B}$, becomes comparable with the average inter-electronic distance. A field of 10 T corresponds to a magnetic length of $\ell_{B}\sim 8 nm$, an order of magnitude longer than the typical interatomic distance in solids. Now, if the carrier density is small enough,  the Fermi surface occupies a tiny portion of the Brillouin zone and the Fermi wavelength of the Bloch electrons extends over many interatomic distances. This is the case of bismuth and graphite, which as bulk semi-metals host a dilute liquid of carriers of both signs.

In bismuth the carrier density of holes is 3$\times$ 10$^{17}$ cm$^{-3}$, which means that there is roughly one itinerant electron per 10$^5$ atom\cite{edelman}. Carrier density in graphite is an order of magnitude larger. But, graphite is a layered material and its Fermi surface consists of very elongated ellipsoids, which extend across the Brillouin zone\cite{brandt}. The cross section of the Fermi surface perpendicular to a magnetic field applied along the high symmetry axis are very comparable in bismuth and graphite. In both cases, when the field is aligned along this axis (the trigonal axis in bismuth and the c-axis in graphite), the quantum limit is attained at a magnetic field lower than 10 T (9 T in bismuth and 7.4 T in graphite).

\section{Nernst quantum oscillations}
\begin{figure}
% Figure 1
\begin{center}
 \resizebox{!}{0.6\textwidth}{\includegraphics{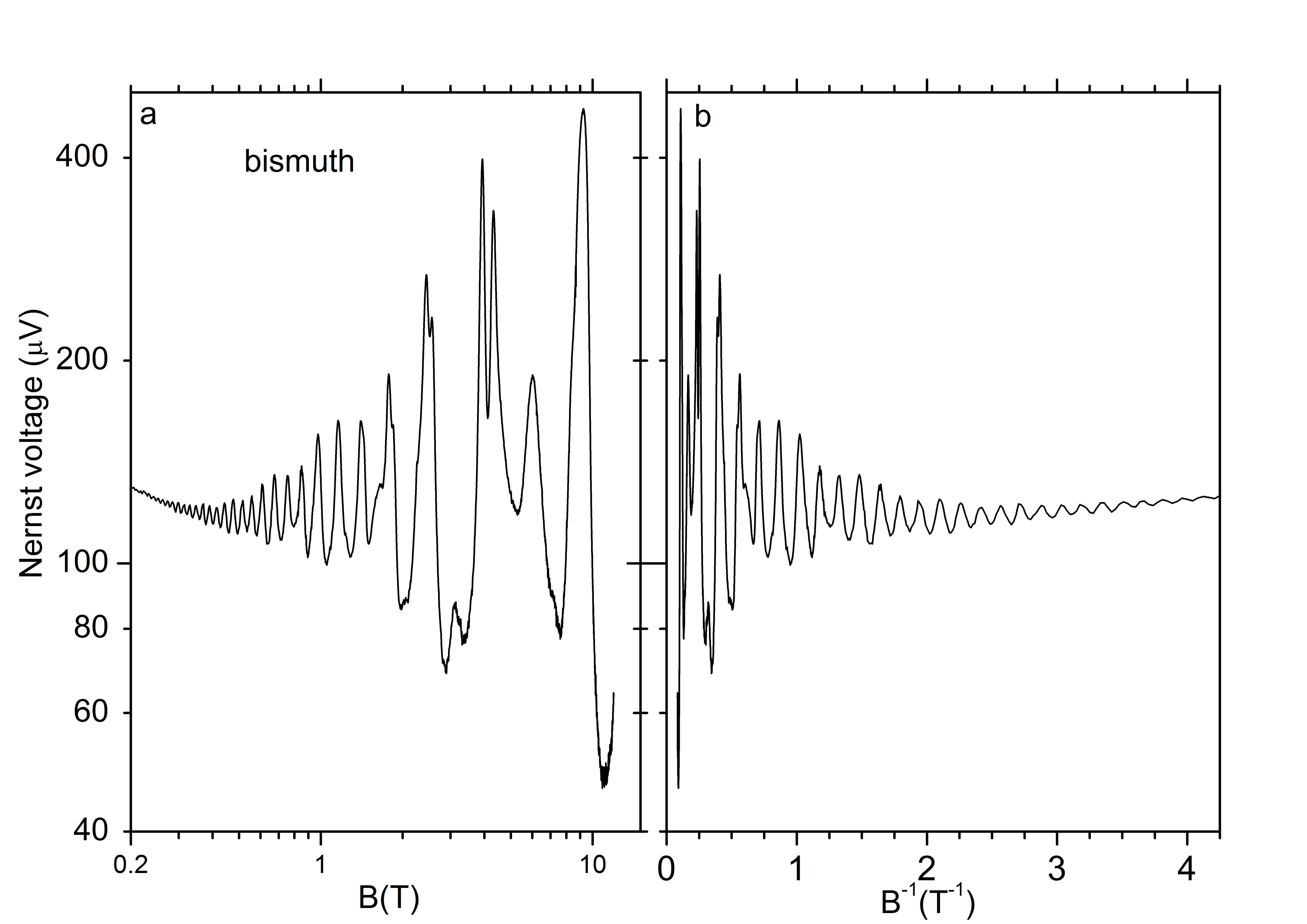}}
\caption{\label{Fig1} Quantum oscillations of the Nernst voltage in bismuth for a field along trigonal axis at T= 0.31 K. The thermal gradient is applied along the bisectrix axis. We present the same data as a function of magnetic field (a) and inverse of magnetic field (b). }
\end{center}
\end{figure}

We measured Nernst and Seebeck effects with a one-heater-two-thermometer set-up. Reversing the magnetic field  allowed us to distinguish between the even (Seebeck) and odd (Nernst) component of the thermoelectric response.

Fig. 1 presents the raw data obtained in a simple Nernst experiment. The evolution of the transverse DC voltage produced by a constant heat current applied along a bismuth single crystal is monitored as the field is swept. Quantum oscillations become visible  in presence of a magnetic field as small as 0.2 T. In this configuration, the Nernst response, like other transport properties\cite{bompadre,fauque1} is dominated by the contribution of the holes. In Fig. 1b, one can see, on the top of the main oscillations, with a period ($\sim$ 0.15 T$^{-1}$) corresponding to the cross section of the hole Fermi surface a slow modulation. The period of this slower oscillation (0.6 T$^{-1}$) points to the presence of a secondary contribution  with a slightly different period, which would generate beating. The cross section of the electron pocket (0.12 T$^{-1}$) \cite{fauque1} is in good agreement with such an explanation.

As the field is swept and the number of the filled Landau levels decreases, the amplitude of the oscillations increase and dominate by far the non-oscillating background. The last and the most prominent peak at 9 T is associated with a six-fold enhancement of the Nernst signal, to be compared with  a mere 30 percent drop in electric resistivity caused by the Shubnikov-de Haas oscillations\cite{fauque1}.

Giant Nernst oscillations with the approach of the quantum limit were observed in graphite with similar features\cite{zhu}.

\section{Indexing Landau levels}

\begin{figure}
% Figure 2
\begin{center}
 \resizebox{!}{0.6\textwidth}{\includegraphics{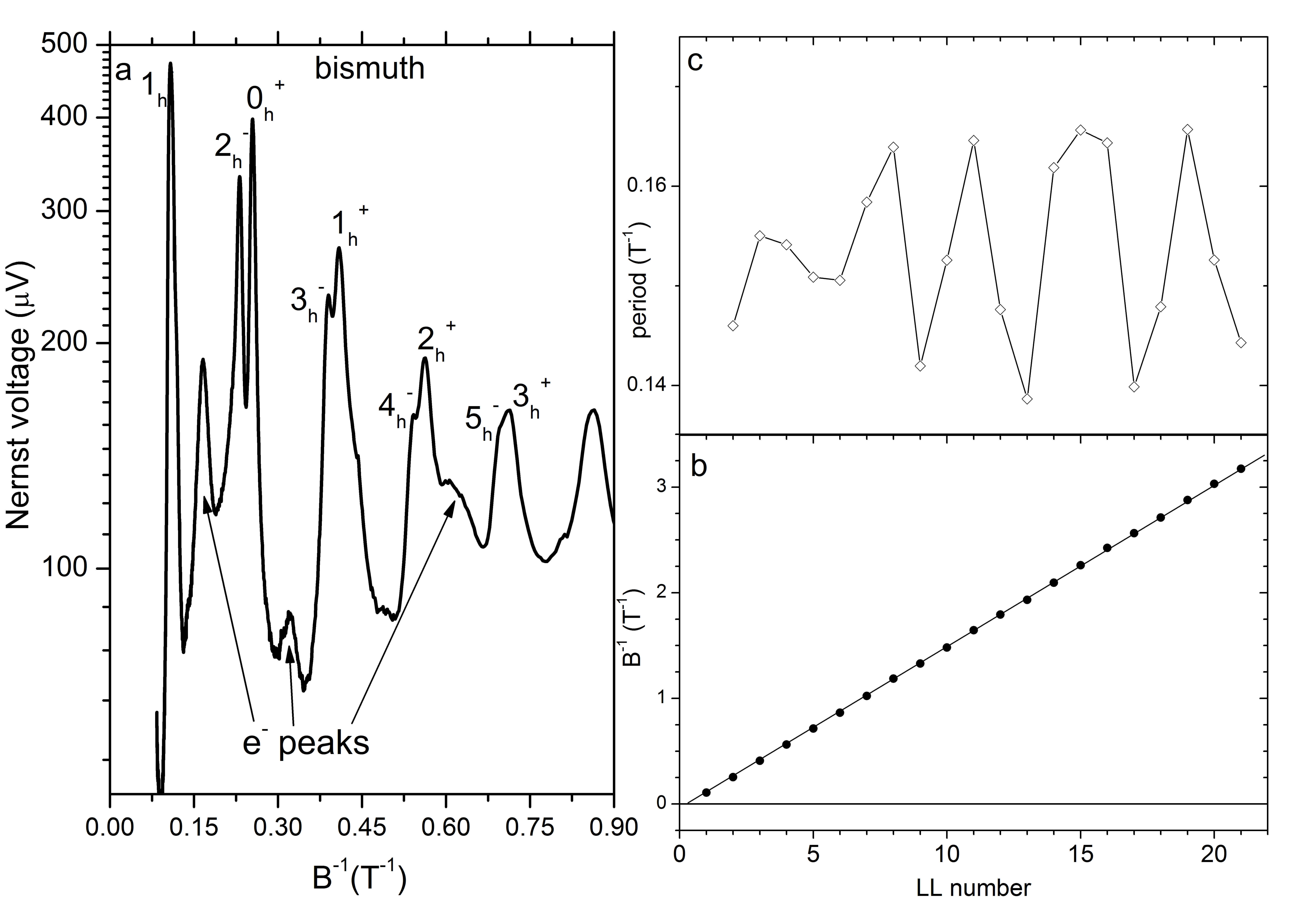}}
\caption{\label{Fig2} a) Nernst signal as a function of the inverse of magnetic field in bismuth. Peaks associated with hole Landau levels are indexed. b) The B$^{-1}$ position of the Nernst peaks vs. their Landau level index. Only those associated with one spin orientation ($^-$) is included. c) The period of oscillations defined as the distance between two adjacent peaks. Ripples are a consequence of the field-induced modification of the Fermi level.}
\end{center}
\end{figure}

With increasing magnetic field, Zeeman energy differentiates carriers of different spins. This can be see in the Nernst data of both bismuth (Fig. 2) and Highly-Oriented Pyrolitic Graphite (HOPG) (Fig. 3), the Nernst peaks associated with the lowest Landau levels become Zeeman-split. One can easily associate a Nernst peak, with a Landau level following  previous studies on bismuth\cite{bompadre} and graphite\cite{woollam}.

Let us note and underline a few differences between the two systems. In graphite, in contrast to bismuth, the Nernst peaks for holes and electrons display comparable amplitudes. This is not very surprising. In graphite,  the electron and the hole ellipsoids composing the Fermi surface, lie parallel, while in bismuth their longer axes are almost perpendicular to each other. Carriers are more mobile when they travel in the plane perpendicular to the longer axis of the ellipsoid. It is also striking that in bismuth, the Zeeman energy of the hole pocket is as large as more than twice the cyclotron energy. This leads to a drastic modification of the hierarchy of the Landau levels, causing the 0$^{+}$ sub-level to occur at a field, which is lower than the 2$^{-}$. In graphite, for both electrons and holes, the Zeeman energy is a mere fraction of the cyclotron energy and thus the distance between spin-polarizes Landau sub-levels remain modest.

As a consequence, there is subtle difference between the two systems as the quantum limit is crossed. In bismuth, when the magnetic field exceeds 9T, there remains a single spin-polarized hole Landau sub-level and two electron Landau sub-levels with different spins. In graphite, on the other hand, both holes and electrons reside on their two lowest spin-split Landau sub-levels.

Panel b in both figures present the position of the Nernst peaks in B$^{-1}$ as a function of the index of the Landau level for bismuth(Fig. 2) and graphite (Fig. 3). Note that there are two distinct types of peaks in graphite and a single one in bismuth. A straight line would imply oscillations which are strictly periodic in B$^{-1}$.  A deviation from strict periodicity points to a field-induced modification of the carrier density\cite{bompadre}.

Panel c presents the variation of periodicity, defined as the distance between two adjacent peaks, as a function of the Landau level number. The observed wobbling is a consequence of the continuous adjustment of the Fermi energy with magnetic field in order to keep charge neutrality between electrons and holes\cite{sugihara}.

\begin{figure}
% Figure 3
\begin{center}
 \resizebox{!}{0.6\textwidth}{\includegraphics{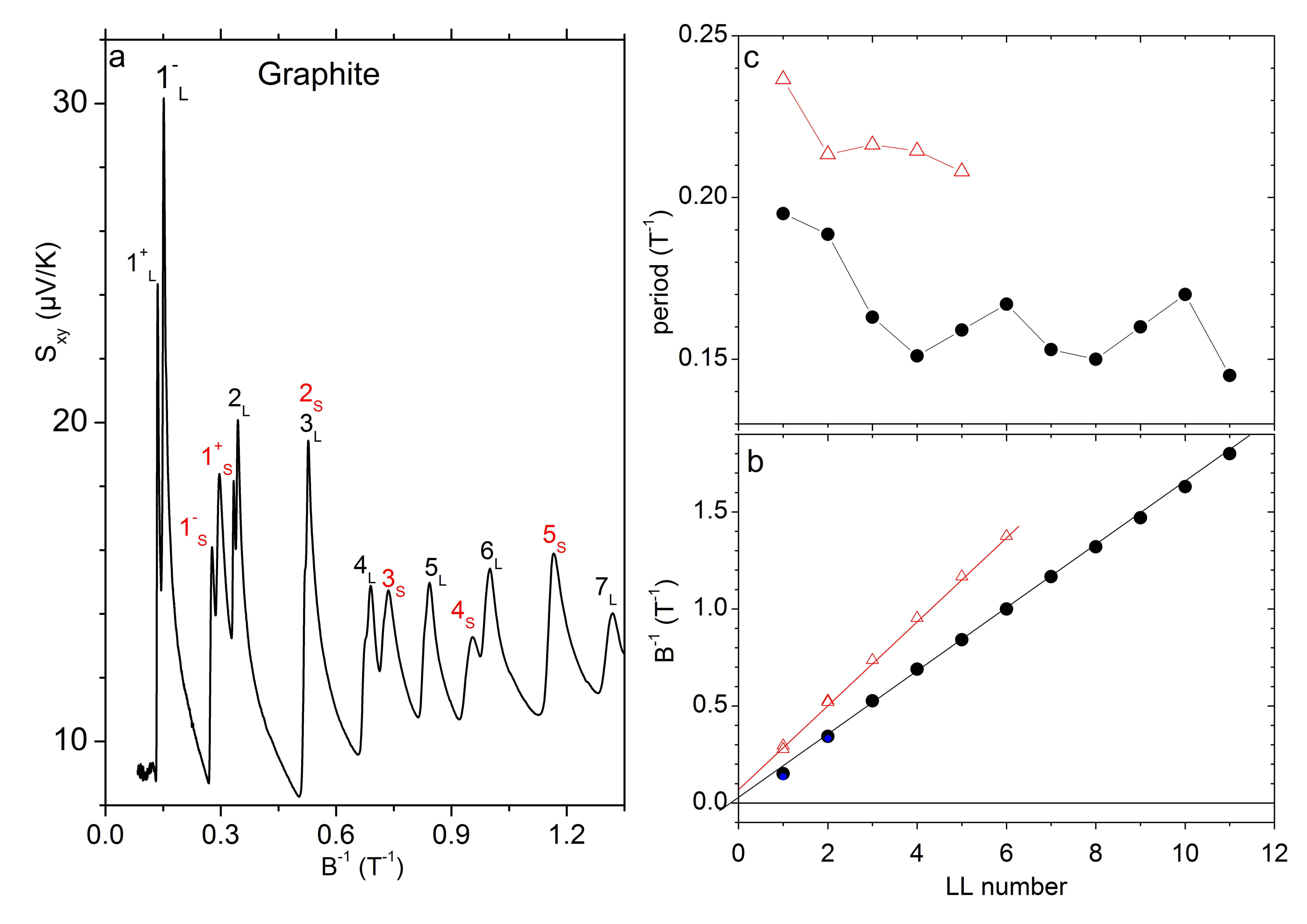}}
\caption{\label{Fig3} a) Quantum oscillations of the Nernst signal in  a HOPG sample as a function of the inverse of the magnetic field at T= 0.34K. Peaks indexed L (S) are associated with the larger (smaller) Fermi surface. The larger Fermi surface is widely believed to be the electron pocket. b) The B$^{-1}$ position of the  Nernst peaks vs. their Landau level index. c) The period of oscillations defined as the distance between two adjacent peaks. Open triangles (solid circles) represent  the S (L) Nernst peaks.}
\end{center}
\end{figure}

\section{Nernst response, dimensionality and  Seebeck response}

\begin{figure}
% Figure 4
\begin{center}
 \resizebox{!}{0.8\textwidth}{\includegraphics{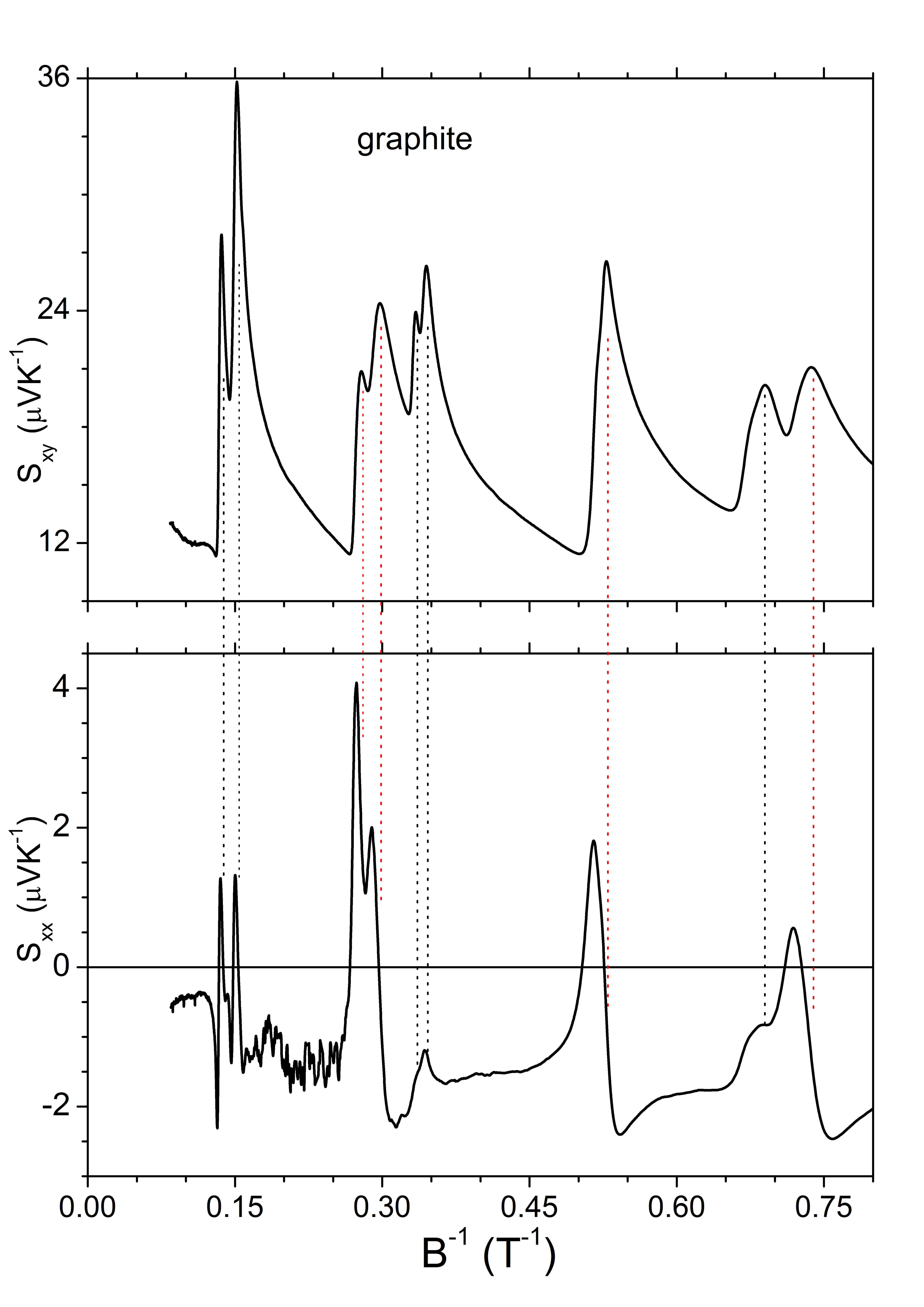}}
\caption{\label{Fig4} The quantum oscillations of the Nernst (a) and Seebeck (b) response obtained simultaneously in graphite at T= 0.55 K. Black (red) dotted vertical lines mark the position of the L (S) Nernst peaks.}
\end{center}
\end{figure}

As seen above, in both bismuth and graphite, the Nernst effect sharply peaks each time a Landau [sub-]level intersects the Fermi level. In three dimensions, this occurs when a Landau tube is squeezed before leaving the Fermi surface. This is concomitant with a van Hove singularity in density of states and has been assimilated to  a Lifshitz topological transition\cite{blanter}.

It is instructive to compare the Nernst response in these bulk semi-metals with what was reported for two-dimensional systems, i.e. graphene \cite{zuev} and semiconducting heterostructures\cite{fletcher}. Such a comparison suggests a striking correlation between the profile of Nernst quantum oscillations and dimensionality. In three-dimensional semi-metals, such as bismuth and graphite, each oscillation consists of a single asymmetric peak. In  two dimensional systems, each Nernst oscillation presents two antisymmetric peaks sandwiching a vanishing signal. This qualitative difference between the 2D and the 3D cases has been one subject of a very recent theoretical analysis\cite{bergman}.

In three dimensions, there is a qualitative difference in the profile of the quantum oscillations of the Nernst and the Seebeck coefficients. Fig. 4 compares the oscillations of the Nernst, S$_{xy}$ , and Seebeck, S$_{xx}$, responses in graphite. As seen in the figure, the Nernst oscillations are not only an order of magnitude larger, but present a much simpler structure. Roughly speaking, maxima in S$_{xy}$ are concomitant with a sign change in S$_{xx}$.  There is no noticeable difference between the profile of the S$_{xy}$ response of hole-like carriers and electron-like carriers. In the case of S$_{xx}$, on the other hand, the anomalies for one type of carriers are weaker than the other. This may be a consequence of the different energy dispersion of the two pockets.

\section{Beyond the quantum limit}

\begin{figure}
% Figure 5
\begin{center}
 \resizebox{!}{0.8\textwidth}{\includegraphics{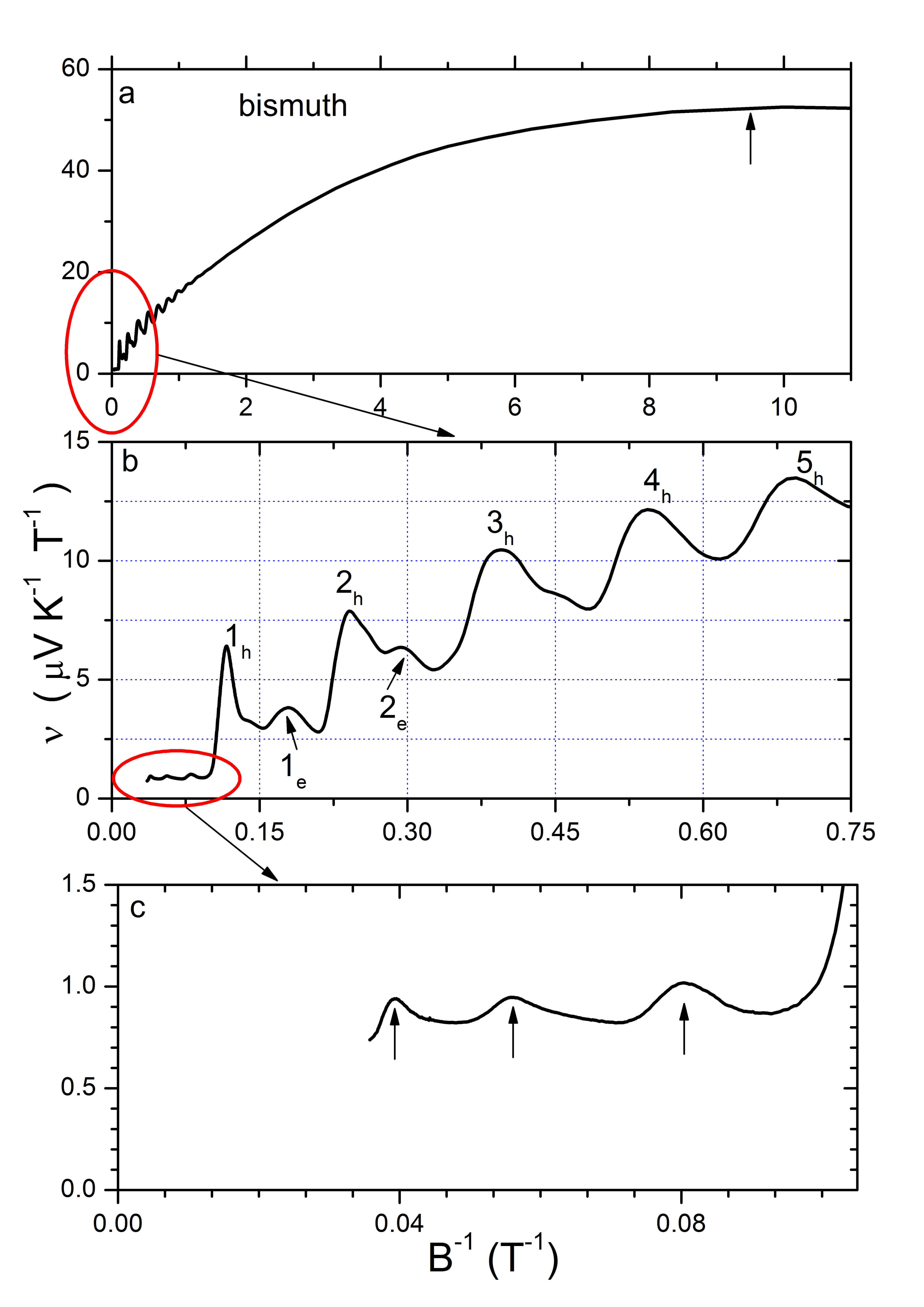}}
\caption{\label{Fig5} a: Quantum oscillations of the Nernst coefficient ($\nu=\frac{S_{xy}}{B}$) in a bismuth single crystal at T= 1.5 K with magnetic field oriented along the trigonal axis. The arrow marks the passage from low-field to high-field regime. b: A zoom to a restricted window in panel a. Quantum oscillations are visible. c: A zoom to a restricted window in b. The three peaks observed beyond the quantum limit cannot be attributed to any known Landau level.}
\end{center}
\end{figure}

The extension of the Nernst measurements to stronger magnetic fields led to the revival, after two decades of hibernation, of another field of investigation: the ground state of a three-dimensional electron gas in an ultra-strong magnetic field\cite{halperin}. Contrary to the two-dimensional case, little is established, theoretically and experimentally, about the fate of a three-dimensional electron gas pushed beyond the quantum limit. In 2007, a study of the Nernst response in bismuth, found unexpected Nernst peaks above 9 T for a field applied along the trigonal axis, which could not be attributed to any known Landau level.

However, following this discovery, two independent set of theoretical calculations found that the one-particle spectrum of bismuth is complex \cite{alicea,sharlai}. According to these calculations, the Landau level crossing of the three electron ellipsoids occur at a field which sharply shifts with angle as the field is slightly tilted off the trigonal axis. The angular dependence of the field scale associated with this crossing was in very good agreement with a jump in magnetization found in angular-dependent torque study\cite{li}. It was suggested\cite{sharlai} that the Nernst anomalies could be attributed to the intersection of Landau levels of the electron ellipsoids assuming a small misalignment of a few degrees. Due to the absence of in situ orientation of the crystal in the initial Nernst experiment\cite{behnia2} this possibility could not be ruled out. Recently, angular-dependent Nernst measurements established that even in the case of perfect alignment between the magnetic field and the trigonal axis there are three unexpected Nernst peaks\cite{yang}.

Fig. 5 presents the variation of the Nernst coefficient ($\nu=\frac{S_{xy}}{B}$) from very low fields to 28 T at T=1.5K. The data are plotted as a function of B$^{-1}$ and the magnetic field is aligned along the trigonal axis with sub-degree accuracy. Three distinct regimes are observable. i) The low-field regime ($\omega_{c} \tau <$ 1), S$_{xy}$ is field linear and $\nu$ is independent of the magnetic field and there are no oscillations. ii) In the high-field regime ($\omega_{c} \tau >$ 1), S$_{xy}$ saturates and  $\nu$ becomes decreases with increasing field. Gradually, quantum oscillations become visible. iii) Above 9 T, beyond the quantum limit, where no further structure in the non-interaction picture is expected, we resolve three additional peaks at about 13 T, 18 T and 26 T. It appears that there are additional Landau tubes in this extreme quantum limit presumably due to electron interaction.

The three additional Nernst peaks have been observed in all six crystals studied by our group. Moreover, as seen in Fig. 6, which presents data for 4 different crystals, they are more prominent in samples with higher RRR (=$\frac{\rho(300 K}{\rho(4.2 K}$), which is often used as a measure of the electron mean-free-path in the bulk of the sample. On the other hand, there is no visible correlation between the amplitude of the peaks and the thickness of the sample.

Both the size of these three additional peaks and their temperature dependence links them to the three electron pockets\cite{yang}. On the other hand, in another Nernst experiment extended up to 45 T, but without \emph{in situ} control of the field orientation, a large peak was resolved around 38 T. The  latter peak is almost as large as the 9 T prominent peaks caused by the crossing of the lowest hole Landau level and suggests that it is associated with the hole-like pocket\cite{fauque2}.

Doping bismuth with antimony leads to a decrease in carrier density and pulls down the quantum limit. Ultraquantum Nernst peaks were also observed beyond the quantum limit in Bi$_{0.96}$Sb$_{0.04}$\cite{banerjee}.

In summary, additional Landau sub-levels (three for electrons and one for holes) are detectable beyond the quantum limit. They remain unexplained in the non-interacting picture and their explanation may require a better understanding of electron interactions in a three-dimensional multi-valley electron system pushed beyond the quantum limit.

\begin{figure}
% Figure 6
\begin{center}
\resizebox{!}{0.6\textwidth}{\includegraphics{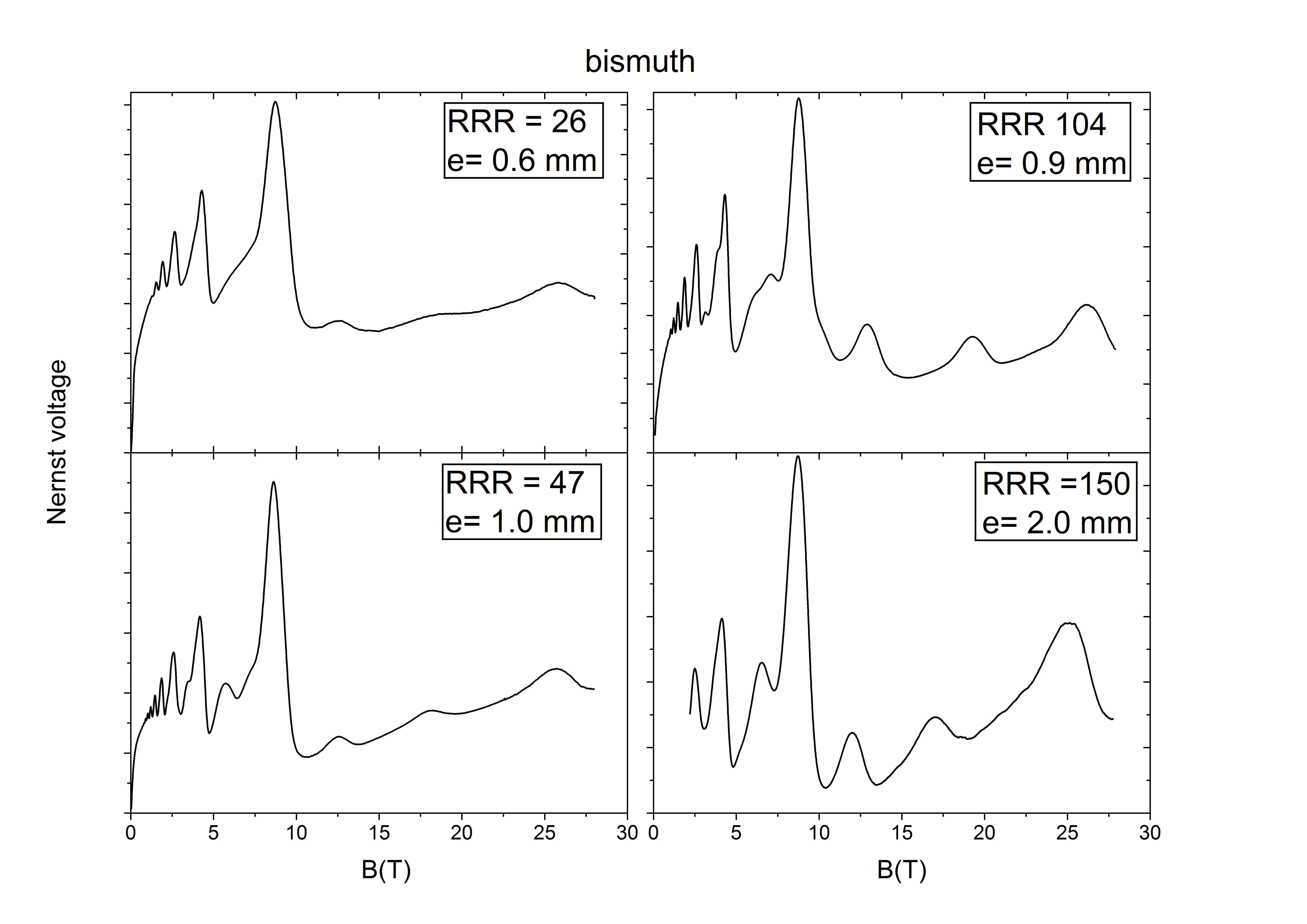}}
\end{center}
\caption{\label{Fig6} Nernst voltage up to 28 T in four bismuth single crystals at T= 1.5 K for field along the trigonal axis. For each sample, the ratio of its room-temperature to helium temperature resistivity (RRR) and its thickness (e) are given. The three ultraquantum peaks are more prominent in cleaner crystals.}
\end{figure}

\section{Acknowledgements}
It is a pleasure to acknowledge  stimulating discussions with J. Alicea, L. Balents, D. Bergman, V. Oganessyan, A. Millis, G. Mikittik, Y. Sharlai and A. Varlamov. This work is supported by Agence Nationale de Recherche (ANR-08-BLAN-0121-02) as part of DELICE and by the Dnipro program.

\end{document}